\documentclass{article}

\usepackage{graphicx}
\usepackage{psfig}
\usepackage{epsfig}
\usepackage[round]{natbib}

\title{Probabilistic forecasts of temperature: measuring the utility of the ensemble spread}

\author{Stephen Jewson\footnote{\emph{Correspondence address}: RMS, 10 Eastcheap, London, EC3M 1AJ, UK.
Email: \texttt{x@stephenjewson.com}}}

\begin{document}

\maketitle

\begin{abstract}
The spread of ensemble weather forecasts contains information
about the spread of possible future weather scenarios. But how much
information does it contain, and how useful is that information in
predicting the probabilities of future temperatures? One
traditional answer to this question is to calculate the
spread-skill correlation. We discuss the spread-skill correlation
and how it interacts with some simple calibration schemes. We then
point out why it is not, in fact, a useful measure for the amount
of information in the ensemble spread, and discuss a number of
other measures that are more useful.
\end{abstract}

\section{Introduction}

Forecasts of the expected surface air temperature
over the next 15 days are readily available from commercial forecast vendors.
The best of these forecasts have been proven to be consistently better than climatology
and such forecasts are widely used within industry.
There is also demand within industry for \emph{probabilistic} forecasts of temperature
i{.}e{.} forecasts that predict the whole distribution of temperatures.
Such forecasts are much more useful than forecasts of the expectation alone
in situations where the ultimate
variables being predicted are a non-linear function of temperature, as is commonly the case.

Probabilistic forecasts of temperature can be made rather easily
from forecasts of the expected temperature
using linear regression.
The parameters of the regression model are derived using past forecasts and
past observations after these forecasts and observations have been converted
to standardized anomalies using the climatological mean and standard deviation.
Probabilistic forecasts made in this way provide a standard against which
forecasts made using more sophisticated methods should be compared, and it turns
out that they are hard to beat (our own attempts to beat regression, which have
more or less failed, are summarised in~\citet{jewson04l}).

Regression-based probabilistic forecasts have a skill that doesn't vary with weather state.
It has been shown, however, that the uncertainty around forecasts of the expectation
\emph{does} vary with weather state and that these variations are predictable, to a certain
extent, using
the spread of ensemble forecasts (see, for example, ~\citet{kalnay}, and many others).
What is not clear is whether the level of predicability
in the variations of the uncertainty is useful in any material sense or whether the beneficial
effect on the final forecast of the temperature distribution is too small to be relevant.
How might we investigate this question of how much useful information there is in the ensemble spread?

One method that is frequently used to assess the amount of information in the spread
from ensemble forecasts is the spread-skill correlation (SSC), defined in a number
of different ways (see for example~\citet{barker91}, \citet{whitaker98} and~\citet{hou01}).
SSC is usually calculated before the ensemble forecast has been calibrated
(i{.}e{.} before it has been turned into a probabilistic forecast).
However, it is the properties of the forecast \emph{after} calibration that we really care about.
In this article we investigate some of the properties of the spread-skill correlation, and in particular
how it interacts with the calibration procedure.
We will show that, under certain combinations of the definition of the SSC and the
calibration procedure, the SSC is the same before and after
the calibration, implying that pre-calibration estimates of the SSC
can be used to predict post-calibration values.

However we also note that even the post-calibration SSC is not a particularly good indicator
of the level of useful information that can be derived from the ensemble spread
and we describe how it can be possible
that the SSC is high but the ensemble spread is effectively useless as a predictor
of the future temperature distribution.

Finally we present some simple measures that improve on the SSC and
that can be used to ascertain whether the information in the ensemble spread
is really useful or not.

\section{The linear anomaly correlation}

We start by reviewing some of the properties of the linear anomaly correlation (LAC).
This will help us understand how to think about the properties of the SSC.

The amount of information in a temperature forecast from an NWP model is commonly measured using the
LAC between the forecast and an analysis.
One of the reasons that the LAC is a useful measure is that it is conserved under linear transformations,
and so if the forecast is calibrated using a linear transformation (such as linear regression)
then the LAC post-calibration is the same as the LAC pre-calibration. This means that
one doesn't actually have
to perform the calibration to know what the post-calibration LAC is going to be.

\section{The spread-skill correlation}

In a similar way the SSC is commonly applied to the output from NWP models to assess the ability of the model
to capture variations in the uncertainty (see for example~\citet{buizza97}).

Four commonly used definitions of SSC are:

\begin{eqnarray}
 \mbox{SSC}_1&=&\mbox{linear correlation}(|e|,s)\\
 \mbox{SSC}_2&=&\mbox{linear correlation}(e^2,s)\\
 \mbox{SSC}_3&=&\mbox{linear correlation}(|e|,s^2)\\
 \mbox{SSC}_4&=&\mbox{linear correlation}(e^2,s^2)
\end{eqnarray}

where $e$ are the forecast errors and $s$ is the ensemble spread.

In the same way that predictions of the mean temperature must be calibrated, so must predictions of the uncertainty.
In~\citet{jewsonbz03a} we argued that both an offset and a scaling are needed in this calibration:
this allows for both the
mean level of uncertainty and the amplitude of the variability of the uncertainty to be set correctly.
We have proposed and tested various models that can be used for this calibration: a summary of our
results is given in~\citet{jewson04l}.
All the models we propose are generalisations of linear regression.
The two models of most relevance to the current discussion are standard deviation and variance
based \emph{spread regression} models defined by:

\begin{equation}\label{sr1}
 T_i \sim N (\mbox{mean}=\alpha+\beta m_i, \mbox{standard deviation}=\gamma+\delta s_i)
\end{equation}

and

\begin{equation}\label{sr2}
 T_i \sim N (\mbox{mean}=\alpha+\beta m_i, \mbox{variance}=\gamma^2+\delta^2 s_i^2)
\end{equation}

where $T_i$ is the temperature anomaly on day $i$,
$m_i$ is the ensemble mean anomaly on day $i$,
$s_i$ is the ensemble spread anomaly on day $i$
and where anomalies are defined by subtracting a climatological seasonal cycle in the
mean and dividing by a climatological seasonal cycle in the spread.
$\alpha, \beta, \gamma$ and $\delta$ are
free parameters:
we call $\gamma$ and $\gamma^2$ the \emph{spread-skill bias correction} and $\delta$ and
$\delta^2$ the
\emph{spread-skill regression coefficient}, while we call $\gamma+\delta \overline{s}$  and
$\gamma^2+\delta^2 \overline{s^2}$ the \emph{spread-skill offset}.

Which of the standard deviation or variance based calibration models is better
is not clear a-priori
but can be answered for any particular data set by comparing the in-sample or out-of-sample
likelihoods achieved by the two models.

It would be very useful if the SSC (for any of the above definitions)
were the same before and after
calibration (for either of the above calibration methods).
Then, as with linear correlation, the pre-calibration SSC could be used to predict
the post-calibration SSC and we would not actually have to perform the calibration
to calculate the post-calibration SSC.

We now investigate whether the SSC has this useful property, which we will call conservation.

\section{Conservation properties of the spread-skill correlation}

The conservation properties of the SSC are straightforward
and somewhat obvious. They can be derived based on the
observation that linear correlations are not affected by linear transformations
of either variable.

Under the standard deviation based spread regression model
the spread skill correlation defined as either SSC$_1$ or SSC$_2$
will be conserved because these measures base the SSC on $s$ and
the calibration of $s$ is simply a linear tranformation.
The SSC measures based on $s^2$ will not, however, be conserved when using
standard deviation based spread regression.

Alternatively under the variance based spread regression model
the spread skill correlation define as either SSC$_3$ or SSC$_4$
will be conserved because these measures base the SSC on $s^2$
and the calibration of $s^2$ is now a linear transformation.
However SSC$_1$ and SSC$_2$ will not be conserved under the variance based spread regression model.

Together these results suggest
that the choice of which SSC measure to choose is not arbitrary but should be influenced
by whichever of the calibration models works better for the data in hand.

\section{The offset problem}

We have shown that the SSC can be conserved during calibration as long as the definition of SSC is
chosen to match the method used for the calibration.
There is, however, a problem with the SSC as a measure for the amount of information
in a probabilistic forecast.
This problem is caused by the spread-skill offset given by $\gamma+\delta \overline{s}$
in equation~\ref{sr1} and by $\gamma^2+\delta^2 \overline{s^2}$ in equation~\ref{sr2}.

When the offset is large relative to the amplitude of the variability of the uncertainty we find ourselves
in a situation in which predictions of the variations of the uncertainty are more or less irrelevant, even
if they are very good, simply because they don't contribute much as a fraction of the total uncertainty.

In such cases the SSC may be large but the ensemble spread could be ignored without reducing the skill of the
calibrated forecast: linear regression would work as well as spread regression.
We clearly need other measures to assess whether the spread is really useful that take into account
the \emph{size} of the calibrated variations in uncertainty.
Since this question depends crucially on the offset
and the offset can only be derived during the calibration procedure
it will not be possible to estimate the usefulness
of the spread before calibration has taken place.

This is a fundamental difference between forecasts of spread and forecasts of the expectation, since, as we
have seen, it \emph{is} possible to estimate the information in a forecast of the expectation before the calibration
has taken place. This difference arises because when we predict the mean temperature we are
concerned with predicting changes from the normal while when we predict the uncertainty we are only interested
in the extent to which our estimate of the uncertainty improves the forecast of the temperature distribution.
Thus we are interested in actual values of the uncertainty rather than just departures from normal.

\section{Other measures of the utility of ensemble spread}

Because of the offset problem with the SSC we now suggest some alternative methods
for measuring the usefulness of the ensemble spread.
All of these measures can only be calculated \emph{after} calibration,
as explained above.

\subsection{Coefficient of variation of spread}

Our first measure is the \emph{coefficient of variation of spread} defined as:

\begin{equation}
\mbox{COVS}=\frac{\sigma_\sigma}{\mu_\sigma}
\end{equation}

where $\sigma_\sigma$ is the standard deviation of variations in the uncertainty or the spread,
and $\mu_\sigma$ is the mean level in the uncertainty or the spread.

COVS was introduced in~\citet{jewsonbz03a} and measures the size of the variations of the
spread relative to the mean spread. Values for the COVS versus lead time for ECMWF ensemble
forecasts for London Heathrow are given in that paper.

If the post-calibration COVS is small then that implies that the variations
in the uncertainty are small relative to the mean uncertainty, and, depending on the
level of accuracy required, that it may be reasonable to ignore
the variations in the uncertainty completely and model it as constant i{.}e{.} that linear
regression may be as good as spread regression.

\subsection{Spread mean variability ratio}

The limitation of using the COVS to understand the importance of
variations in the ensemble spread is it doesn't take into account the size of the
variations in the mean temperature.
One can imagine the following two limiting cases:

\begin{enumerate}

    \item The expected temperature is the same every day but the standard deviation
    of possible temperatures varies. In this case forecasts of the uncertainty
    of temperature would be very useful. We call this a `\emph{mean constant spread varies}' world.

    \item The expected temperature varies from day to day but the standard
    deviation of possible temperatures is constant. In this case forecasts
    of the uncertainty of temperature would not be useful.
    We call this a `\emph{mean varies spread constant}' world.

\end{enumerate}

In order to distinguish between these two scenarios we define the
\emph{spread-mean variability ratio} as:

\begin{equation}
 \mbox{SMVR}_1=\frac{\sigma_\sigma}{\sigma_\mu}
\end{equation}

where $\sigma_\sigma$ is the standard deviation of variations in the uncertainty or the spread
and $\sigma_\mu$ is the standard deviation of variations in the expected temperature.

An alternative definition based on variance would be:
\begin{equation}
 \mbox{SMVR}_2=\frac{\sigma^2_\sigma}{\sigma^2_\mu}
\end{equation}

The SMVR measures the size of variations of the spread relative to the size of the variations of the mean.
Small values of the SMVR imply that we are close to the mean-varies-spread-constant world while large
values of SMVR imply that we are close to the mean-constant-spread-varies world.

\begin{figure}[!htb]
  \begin{center}
   \includegraphics[scale=0.7,angle=0]{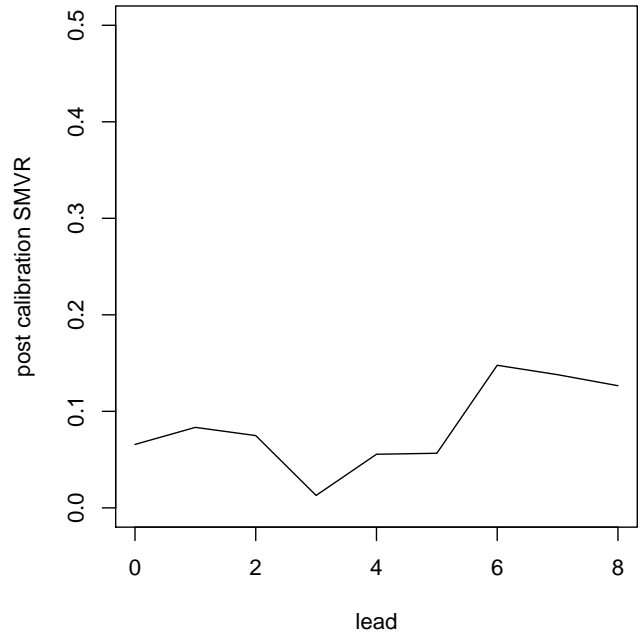}
  \end{center}
  \caption{
The SMVR$_1$ calculated from one year of ECMWF ensemble forecasts
for London Heathrow calibrated using the standard deviation based
spread regression model.
  }
  \label{f:f1}
\end{figure}

Figure~\ref{f:f1} shows the post-calibration SMVR$_1$ for the forecasts used in~\citet{jewsonbz03a}.
We see that the SMVR is small at all leads, with smallest values at the shortest leads.
We thus see that we are much closer to the mean-varies-spread-constant world than we are to the
mean-constant-spread-varies world, and hence that predicting variations in the uncertainty
is likely to be less useful than it would be in a world in which the SMVR  were larger.

%
%
%
%
%
%

\subsection{Impact on the log-likelihood}

The final measure of the utility of forecasts of spread that we present is simply the change
in the cost function that is being used to calibrate and evaluate the forecast. We ourselves prefer
to evaluate probabilistic forecasts of temperature using the log-likelihood from classical
statistics (\citet{fisher1912}, \citet{jewson03d})
and hence we consider the change in the log-likelihood due to the
inclusion of information from the ensemble spread as a measure of how useful that information is.
When we evaluated
the usefulness of the spread in temperature forecasts derived from the ECMWF ensemble using
this method we found that the spread was not very important~\citep{jewson04l}.

One aspect of our comparison of forecasts using log-likelihoods in~\citet{jewson04l} is that
we calculated log-likelihood based on the whole
distribution of future temperatures. This was deliberate: it is predicting the whole distribution of
temperature that we are interested in.
However, if instead we were mainly interested in the tails of the
distribution then a version of the log-likelihood based only on the tails
would be more appropriate and the ensemble spread would perhaps be more useful.

\section{Summary}

We have considered how to measure the importance of variations in the ensemble spread when making probabilistic
temperature forecasts.
First we have considered the interaction between measures of the spread-skill correlation (SSC) and the
methods used to calibrate the forecast. We find that certain definitions of SSC are conserved through the
calibration process for certain calibration algorithms, implying that the choice of SSC measure to
be used should be linked to the choice of calibration method.

However we also discuss why the SSC is not a particularly useful measure of the information in the ensemble
spread and explain how a high value of the SSC does not necessarily
mean that the spread improves the quality of the final forecast because of the possibility of a large
offset in the calibrated uncertainty.

We have discussed some alternative and preferable diagnostics that focus on the role the spread plays in the final
calibrated forecast.
The first of these diagnostics measures the size of variations
in the uncertainty relative to the mean uncertainty
and the second measures the size of variations in the uncertainty relative to the size of the variations in the
expected temperature.
We calculate the latter for a year of forecast data and find that we are much closer to a world
in which the mean varies and the spread is fixed than we are to a world in which the
the spread varies and the mean is fixed. This seems to partly explain why we see so little improvement
in the skill of probabilistic forecasts when we add the ensemble spread as an extra predictor.

\section{Acknowledgements}

Thanks to Jeremy Penzer and Christine Ziehmann for some interesting discussions on this topic.

\section{Legal statement}

SJ was employed by RMS at the time that this article was written.

However, neither the research behind this article nor the writing
of this article were in the course of his employment, (where 'in
the course of their employment' is within the meaning of the
Copyright, Designs and Patents Act 1988, Section 11), nor were
they in the course of his normal duties, or in the course of
duties falling outside his normal duties but specifically assigned
to him (where 'in the course of his normal duties' and 'in the
course of duties falling outside his normal duties' are within the
meanings of the Patents Act 1977, Section 39). Furthermore the
article does not contain any proprietary information or trade
secrets of RMS. As a result, the author is the owner of all the
intellectual property rights (including, but not limited to,
copyright, moral rights, design rights and rights to inventions)
associated with and arising from this article. The author reserves
all these rights. No-one may reproduce, store or transmit, in any
form or by any means, any part of this article without the
author's prior written permission. The moral rights of the author
have been asserted.

The contents of this article reflect the author's personal
opinions at the point in time at which this article was submitted
for publication. However, by the very nature of ongoing research,
they do not necessarily reflect the author's current opinions. In
addition, they do not necessarily reflect the opinions of the
author's employer.

\clearpage

\bibliography{jewson}

\end{document}